# Determination of the Boltzmann Constant Using the Differential – Cylindrical Procedure

X J Feng[1], J T Zhang[1*], H Lin[1], K A Gillis[2], M R Moldover[2*]

[1] National Institute of Metrology, Beijing 100013, China

[2] National Institute of Standards and Technology, Gaithersburg, MD 20899-8360，USA

**Abstract**

We report in this paper the progresses on the determination of the Boltzmann constant $k_B$ using the acoustic gas thermometer (AGT) of fixed-length cylindrical cavities. First, we present the comparison of the molar masses of pure argon gases through comparing speeds of sound of gases. The procedure is independent from the methodology by Gas Chromatography – Mass Spectrometry (GC – MS). The experimental results show good agreement between both methods. The comparison offers an independent inspection of the analytical results by GC-MS. Second, we present the principle of the novel differential-cylindrical procedure based on the AGT of two fixed – length cavities. The deletion mechanism for some major perturbations is analyzed for the new procedure. The experimental results of the differential – cylindrical procedure demonstrate some major improvements on the first, second acoustic and third virial coefficients, and the excess half-widths. The three acoustic virial coefficients agree well with the stated-of-the-art experimental and theoretical (the *ab initio* calculation) results. The new method is characterized by much less correction-dependence. The differential – cylindrical procedure results in the preliminary determination of $k_B$ of $1.380\ 650\ 6\times10^{-23}$ J·K$^{-1}$ with the relative standard uncertainty of $2.3\times10^{-6}$. The new value is in $1.3\times10^{-6}$ above the adjusted Boltzmann constant given in CODATA 2010. The resultant gas constant $R = k_B N_A$=8.314 473 0 J·mol$^{-1}$·K$^{-1}$ with the same the relative standard uncertainty.

---

[*] Correspondent author     email: zhangjint@nim.ac.cn    tel. +8610 6452 6172

[*] Correspondent author     email: michael.moldover@nist.gov

## 1. Introduction

The Boltzmann constant $k_B$ relates the thermodynamic temperature to the kinetic energy of particles. For an equilibrium ideal gas system, the Boltzmann distribution leads that the kinetic energy of ideal gas of mass $m$ is measured by the thermodynamic temperature, $(1/2)mv_{RMS}^2 = (3/2)k_B T$. The kinetic energy connects to the zero-frequency speed of sound $c_0$ and the heat-capacity ratio $C_p^0/C_v^0 \equiv \gamma_0$ of gas, $v_{RMS}^2 = (3/\gamma_0)c_0^2$. For a monatomic gas $\gamma_0$ is exactly 5/3. The state-of-the-art acoustic resonance method underpins the high accurate determination of $c_0$ as well as $k_B$ since the 1970s [1,2,3,4,5,6,7,8,9,10]. For ideal gas, the Boltzmann constant is obtained by:

$$k_B = c_0^2 M / (T \gamma_0 N_A) \quad (1)$$

where $M$ is the molar mass of gas; $N_A$ is the Avogadro constant with the relative standard uncertainty of $3 \times 10^{-8}$ [11].

In general, the molar mass of argon used for acoustic resonant experiment is calculated by knowing the isotopic abundance ratios of Ar40/Ar36 and Ar40/Ar38. The ratios are measured by Gas Chromatography–Mass Spectrometry (GC-MS) systems. The sensitivity and the discrimination factors of GC-MS have to be calibrated by reference argon mixture of known isotopic ratios produced by the gravimetric method [12]. The BIP argon sample used in our previous study [9] had been analyzed by the GC-MS system in the Center for Gas Metrology, Korea Institute of Standards and Science (KRISS) for the isotopic abundance ratios of Ar40/Ar36 and Ar40/Ar38. After, the BIP sample from the same cylinder [9] and the BIP Plus sample used in the latest experiment [10] were analyzed together by the GC-MS system in the state Key Laboratory of Petroleum Resource Research, Chinese Academy of Sciences (KLPRR CAS) for the ratios of Ar40/Ar36 and Ar40/Ar38. Thus, the values given by the second analysis can be linked to the gravimetric primary standard of KRISS through the BIP sample.

Eq. (1) shows that the molar masses of different ideal gases could be compared through the measurements of speeds of sound on an isotherm. This idea had been practiced in the early study [3]. The comparison implies the assumption that the molar mass of the sample gas from a certain container has to be a constant. Such a comparison offers an independent inspection for analytical results with GC-MS. We have the BIP and the BIP Plus samples analyzed respectively by KRISS and KLPRR with GC-MS. We are motivated by the idea to conduct acoustic comparison of the molar masses for those samples and another BIP Plus sample.

According to Ref. [13], cavities of any shapes can be used for forming acoustic resonances. For the highest possible accuracy, acoustic resonance measurements in dilute gases are implemented in the radially symmetric modes with spherical or quasi-spherical cavities and the longitudinal and radial modes with cylindrical cavities. Heretofore, the determinations of $k_B$ relying on the longitudinal non-degenerate modes have been only reported by the studies with the variant length cylindrical cavity [1,2] and with the fixed-length cylindrical cavities [9,10]. The fundamentals of the fixed-length cylindrical method have been detailed in the previous studies by the authors [9,10,14,15,16,17,18]. The method is characterized by the following points. (1)

Gas is enclosed steadily in cavity so to avoid the effect of gas contamination and injection. Measurements consume minor amount of gases so to ease the sampling of gases for the isotopic and the impurity analyses. (2) Driving and detecting transducers, piezoelectric transducers (PZT), have diaphragms separating the cavity from the PZTs and gas pressure vessel. The diaphragms maintain the integrity of the cavity wall. The PZT detector having much large capacitance allows the application of coaxial cable-connection remote amplifier. (3) Cylindrical cavity is mechanically robust and geometrically simple that the assembling of resonator is convenient without relying on delicate and complex technique. (4) The quality factors of cylindrical cavities are significantly lower than those of spherical or quasi-spherical cavities so that the random noise is significant at low gas pressures. (5) The endplate bending and the free recoil of cavity are extremely hard to be well characterized. The admittances of PZT transducers and gas fill ducts were characterized in our previous studies [9,10,14,15]. Nevertheless, correcting the perturbations caused by those imperfections to the level of part of million is still considerably difficult.

AGT of fixed-length cylindrical cavity probes a resonant frequency by scanning nearby the certain mode. The method is challenged by the difficulties extracting unperturbed resonant frequencies from perturbed measurements. The extracting accompanies with a number of corrections of perturbations caused by imperfection effects. Some imperfections, such as the endplate bending, are extremely hard to be well characterized. AGT of variant-length cylindrical cavity is characterized by its way probing a resonance by varying cylindrical cavity length in a constant acoustic frequency. Because the some perturbations are frequency – dependent, a constant frequency makes the imperfections have equal acoustic admittances. The deletion mechanism resorting to differential procedure is then applicable instead of corrections of some major perturbations. Besides, the displacement of a movable endplate is counted instead of absolute measurement of the cylinder length. This eases the length measurement. Nevertheless, AGT of variant-length cylindrical cavity has some significant drawbacks. The cavities have to be slim in order for sufficient moving distances and stiffness of the moving piston endplate. There is great challenge for aligning the piston and maintaining a constant angle between the axes of the cylindrical cavity and the piston.

AGT of fixed – length cylindrical cavity inherits the advantage of simplicity in machining and installation, and the disadvantage of unequal admittances between different modes. Coupling the advantages of both cylindrical methods would yield some preferring properties for the cylindrical acoustic thermometry. On this consideration, the authors have proposed the novel differential – cylindrical procedure based on a pair of fixed – length cylindrical cavities. The core of the procedure is to obtain equal admittance for the pair cavities so that the deletion mechanism is applied by the differential procedure. We presented in this paper the principle of the differential procedure using a pair of fixed-length cylindrical cavities. We demonstrated experimentally the deletion effectiveness with the new AGT operating in the differential procedure. We conducted the comparisons to show that the new procedure improves the measurements of the square speeds of sound, the second and third acoustic coefficients and the half-widths of resonances. The new determination of $k_B$ is obtained with the differential procedure.

## 2. Molar Mass Comparison

### 2.1 Molar mass depending on analysis of GC-MS

BIP argon produced by the Air Products was applied in our first determination of $k_B$ [9]. A sample of the gas collected downstream to the getter was analyzed for the isotopic ratios of Ar40/Ar36 and Ar38/Ar36 by the Gas Chromatography–Mass Spectrometry (GC-MS) system (Finnigan MAT271) in the Center for Gas Metrology, Korea Institute of Standards and Science (KRISS). The sensitivity and the discrimination factors of the spectrometer were calibrated by the reference argon mixture of known isotopic ratios produced by the gravimetric method [12]. The procedure is an absolute method that had been used in Ref. [19,20,21]. The analysis determines the value $M_{Ar}$ = (39.947 843 ± 0.000 028) g·mol$^{-1}$ that was calculated from the measured isotopic ratios Ar38/Ar36 = 0.1892±0.00035 and Ar40/Ar36 = 299.59±0.31 with the relative uncertainty, $u_r(M_{Ar})$ = 0.7×10$^{-6}$ (1σ).

We used two different grades of argon, the BIP and the BIP Plus in our latest determination of $k_B$ [10]. The BIP argon was supplied from the original gas cylinder utilized by the first determination of $k_B$ [9]. Thus, the BIP argon has the specific isotopic ratios analyzed in KRISS. We conducted the analysis of the isotopic ratios of the BIP Plus argon by the GC – MS system MAT271 in the state Key Laboratory of Petroleum

Resource Research, Chinese Academy of Sciences (KLPRR CAS). Three samples, one of the BIP and two of the BIP Plus argon from the same gas cylinder, were collected at the downstream to the getter of the gas manifold of the cylindrical AGT. Two BIP Plus samples from the same cylinder were labeled in Sample A and B, respectively. The BIP sample was labeled in Sample C. The sampling and labeling were blind for KLPRR CAS. The air was taken as the reference in the analyses by MAT 271 in KLPRR CAS. As stated above, Sample C had the known isotopic ratios. Therefore, Sample C bridged a linkage of the analysis in KLPRR CAS to the primary gravimetric standard of KRISS, having assumed the constant isotopic ratios of Sample C. The samples were analyzed in the order "C-B-A-C-A-B" for the purpose minimizing bias. Each analysis for each sample composed of more than five repeated measurements. The resulting isotopic ratios of each sample were the averages of the repeated measurements. The two runs of analyses for Sample C turn out the inconsistency of $0.016 \times 10^{-6}$ between two averages of the molar masses of Sample C. The inconsistency accounts for the measurement instability of the GC-MS and the inconsistency of the argon isotopic ratios of the local air samples. The repeated measurements in each run yields the non-repeatability bound in $0.26 \times 10^{-6}$. The inconsistency and the non-repeatability were counted in the uncertainty for the linkage to the KRISS primary standard. The analytical results are tabulated in Table 1. The inconsistencies and non-repeatabilities associating with Sample A and Sample B are counted in the uncertainties for the calculation of the molar masses of Sample A and B shown in Table 1. The fractional difference of the molar masses between Sample A and B was $0.04 \times 10^{-6}$. As stated above, Sample A and B were collected from same cylinder. The value of the molar mass $M_{Ar}$ for the BIP Plus was the average of the lumped data of Sample A and B in two runs. The value of $M_{Ar}$ of Sample C is larger than the average of Samples A and B by the fraction $0.82\times10^{-6}$. We summarize the analyses by KRISS and KLPRR in Table 1.

Table 1 Summary of the analyses by KRISS and KLPRR

|   | $M_{Ar}$ (g·mol$^{-1}$) | Non-repeatability ×10$^6$ | Inconsistency ×10$^6$ | Uncertainty* 10$^6$ (%) |
|---|---|---|---|---|
| [1] Sample C | (39.947 843) | 0.12 | 0.08 | (0.70) |
| [2] Sample C | (39.947 843) | 0.15 | | |
| [1] Sample A | 39.947 798 | 0.14 | 0.30 | 0.78 |
| [2] Sample A | 39.947 822 | 0.14 | | |
| [1] Sample B | 39.947 806 | 0.20 | 0.12 | |
| [2] Sample B | 39.947 815 | 0.20 | | |

Note: 1) the superscript 1 and 2 denotes respectively Run 1 and 2

2) the value bracketed is $M_{Ar}$ given by the analysis at KRISS

3) the superscript * denotes the combined uncertainty (1σ) with the linkage to the primary standard of KRISS

**2.2 Comparison of molar masses by AGT**

For dilute gases, there exists the acoustic virial equation:

$$c^2(p,T) - A_3 p^3 = A_0 + A_1 p + A_2 p^2 + A_{-1} p^{-1} \tag{2}$$

in which, the intercept $A_0$ results in the square speed of sound at the ideal gas state, $A_0 \equiv c_0^2 = \gamma_0 k_B T N_A / M$ and $\gamma_0 = 5/3$ for monoatomic argon; $A_1$ and $A_2$ denotes the second and the third acoustic virial coefficients; $A_{-1}$ is an adjustable coefficient accounting for the thermal and dynamic effects of molecule slipping upon cavity wall; $A_3$ is a fixed-value parameter; $c^2$ denotes the square speed of sound of practical gas; $p$ and $T$ stands for gas pressure and temperature, respectively. The surface fitting of Eq. (2) yields the parameters $A_0$, $A_1$ and $A_2$. Both Eq.(1) and (2) feature the theoretical foundation for the determination of $k_B$ with AGTs.

Having fixed $p$ and $T$, Eq. (2) enables mode – to – mode comparisons of the molar masses $M$ of monoatomic gases upon the measurements of $c^2$, that is

$$\frac{M_1}{M_2} = \frac{c_2^2}{c_1^2} \qquad (3)$$

in which, the subscript 1 and 2 stands for monoatomic gas 1 and 2. According to Eq. (3), a comparison was implemented for the molar masses of Sample C, Sample AB from the cylinder of Sample A and B, and Sample D that is a BIP Plus sample from an arbitrary cylinder. The molar mass of Sample C bridges a linkage to the primary standard of KRISS.

We pinned the comparison at 300 kPa and 273.16 K because the AGT of the single fixed – length cavity has been observed of the minimum measurement inconsistency at that pressure. Mode (200), (300) and (400) were applied for the comparison. As we stated in our previous publication [9,10], a getter was upstream to the cavity in the gas manifold in order to diminish the reactive impurity components. For pure argon samples of the BIP and the BIP Plus grades, the getter is claimed to diminish the reactive impurities to less than 0.01 part of million. The natural concentrations of the noble gas impurities in the BIP and the BIP Plus argon had been studied in Ref. [6]. The study inferred the amount fractions of Kr and Xe present bound in $0.01 \times \times 10^{-6}$, the level of neon in argon below $0.03 \times 10^{-6}$, and the limit of $0.01 \times 10^{-6}$ helium in argon. The largest effect on measurements of square speeds of sound of argon is bound in the level of ±

$0.03×10^{-6}$. Accordingly, the effect of the noble gases in argon will be omitted from the comparison.

The comparison was implemented with the previous used AGT of the cylindrical cavity of 80 mm long [10]. More than ten runs were carried out for each sample. The non-repeatabilities for each mode of each sample are given in Table 2. Figure 1 pictures the dispersion of the square speeds of sound of Sample C.

Table 2 Non-repeatability of measurements for Sample C, AB and D

| Sample | Numbers of runs | Non-repeatability of measurements | | |
|---|---|---|---|---|
| | | Mode (200) | Mode (300) | Mode (400) |
| Sample C | 13 | 0.33 | 0.18 | 0.15 |
| Sample AB | 8 | 0.46 | 0.43 | 0.43 |
| Sample D | 25 | 0.40 | 0.31 | 0.29 |

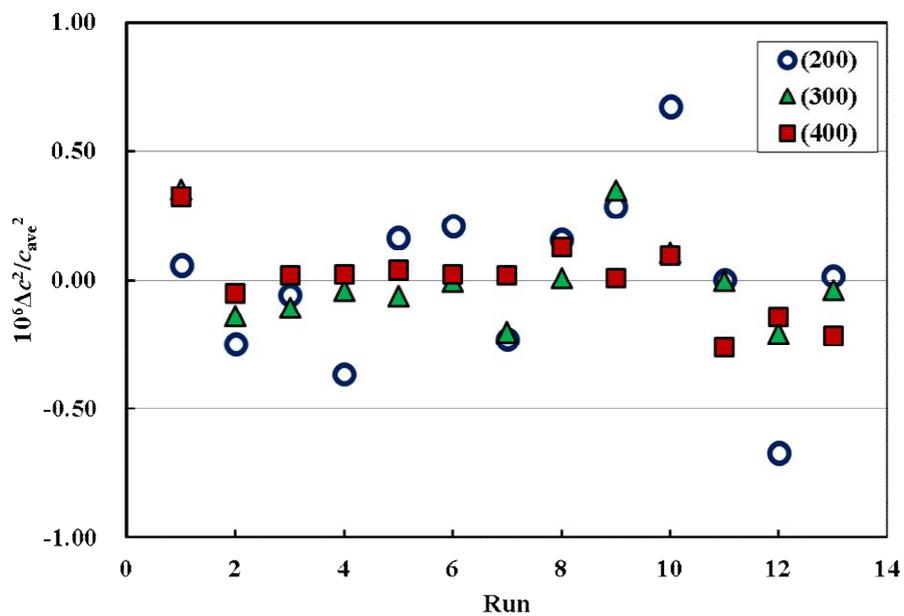

Figure 1 dispersion of the square speeds of sound of Sample C

The measurements of mode (200) appear in a slightly larger dispersion than those of mode (300) and (400). According to Eq. (3), the ratios of molar masses were calculated from mode to mode of each sample with respect to Sample C. The average of the lumped speeds of sound over the entire runs of each mode was calculated for the mode-to-mode comparison. The comparisons with mode (200), (300) and (400) resulted in three ratios that have fractional differences less than $0.1\times10^{-6}$. The average of three ratios was calculated for one sample. Figure 2 diagrams the fractional difference of the molar masses of Sample A, Sample B, Sample AB and Sample D with respect to Sample C.

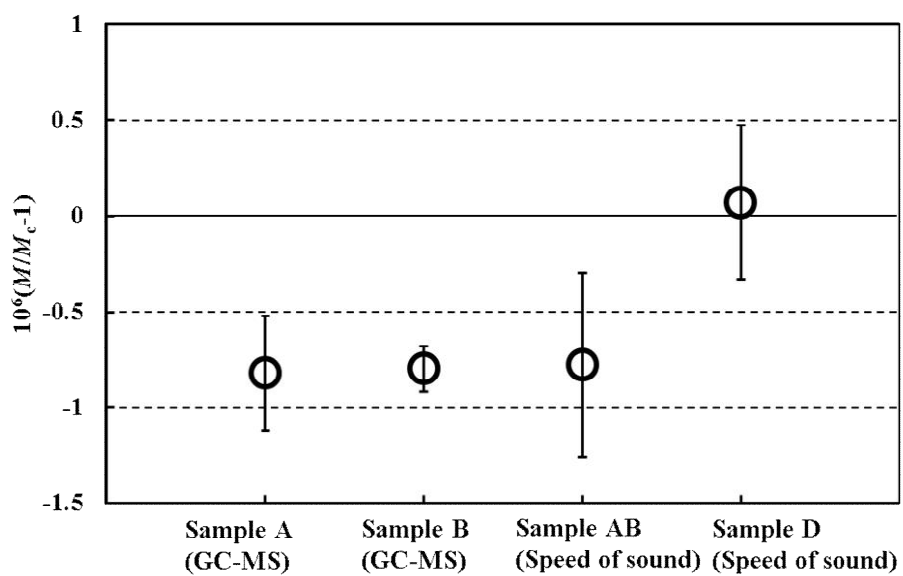

Figure 2 The fractional difference of the molar mass of samples A/B/D relative to that of sample C (uncertainty bar stands for the standard deviation of measurements; the uncertainty of Sample C was not counted in; $M_c$ denotes the molar mass of Sample C)

Figure 2 shows that the results of the GC-MS and the acoustic comparison agree well for the fractional differences of the molar masses of Sample A, B and AB. As we stated above, Sample A, B and AB were collected from the same cylinder. The result of the acoustic comparison agrees well with that of the analyses done by KLPRR CAS with the GC-MS. The acoustic comparison informs that the molar masses of Sample AB and Sample D differ in the fractional difference of $0.8\times10^{-6}$.

### 3. Principle of the Differential-Cylindrical Procedure

Similar to AGTs of spherical and quasi-spherical cavities, AGTs of fixed-length cylindrical cavities are relied on the absolute measurements of cavity dimensions and resonant frequencies. The procedure is technologically challenged by extracting unperturbed resonant frequencies from measured perturbed resonant frequencies through corrections of a number of perturbations caused by imperfection factors. We have catalogued for AGTs of fixed – length cavities in Ref. [9,10] four types of imperfection effects, such as gas viscosity, shell motions of cavity shell and free recoil of cavity body, non-zero acoustic admittance of transducers, gas fill ducts. The caused perturbations have to be characterized and corrected from measured perturbed resonant frequencies to extract "unperturbed" values. The characterization for endplate bending is most challengeable because of poor quantitative information on the clamping force applied between the endplates and the cylindrical shell. The caused perturbation is recognized the predominant challenge for AGTs of single fixed-length cylindrical cavity to achieve the level of measurement relative standard uncertainty of $(1 \sim 2)\times10^{-6}$.

We have previously proposed the novel concept of the "two–cylinder" method to combine the advantages of both the fixed and the variant length cylindrical methods [16,18]. In order to highlight the character of the deletion mechanism for perturbations, we name the "two-cylinder" method in this paper the "differential-cylindrical procedure". In the following section, we reduce the principle of the differential – cylindrical method and present analytically the deletion mechanism for some typical perturbations. According to the fundamental acoustic theory [22], a non-zero complex admittance $y(\mathbf{r}_s, f_l)$ acting on a cavity of volume $V$ inside an acoustic resonance formed will yield a complex perturbation on the ideal wavenumber $k_l$, that is:

$$\frac{K_l^2 - k_l^2}{k_l^2} = \left(\frac{\omega}{ck_l^2}\right)\left(\frac{j}{V\Lambda_l}\right)\iint_S y(\mathbf{r}_s, f_l)|\phi_l(\mathbf{r}_s)|^2 \, dS \tag{4}$$

in which, $c$ denotes the speed of sound; $K_l$ denotes the complex wavenumber, $K_l = 2\pi(f_l + jg_l)/c$, in which $g_l$ stands for the half-width of the resonant profile of a perturbed resonant frequency $f_l$; $k_l$ is the ideal wavenumber, $k_l = 2\pi f_l^0/c$ that $f_l^0$ stands for the unperturbed resonant frequency; the volume $V = \pi r^2 L$, that $r$ and $L$ marks the inner radius and the length of the of cylindrical cavity, respectively; $S$ labels the area on which the non-zero admittance $y(\mathbf{r}_s, f_l)$ acts, where $\mathbf{r}_s$ is the vector at the position the admittance $y(\mathbf{r}_s, f_l)$ acting; the circular frequency $\omega = 2\pi f_l$; $\phi_l$ denotes the eigenfuction that is dependent on the ratio of the axial coordinate to the cavity length, and $\Lambda_l = \frac{1}{V}\int(\phi_l)^2 dV$.

The complex quantity $K_l$ can be expanded into the polynomial. Having the polynomial of Eq.(4) cut into the 1$^{st}$ order approximation, the real part is

$$\frac{\Delta f_l}{f_l} \cong -\frac{1}{2}\text{Re}\left(\left(\frac{c \cdot j}{2\pi f_l V \Lambda_l}\right)\iint_S y(\mathbf{r}_s, f_l)|\phi_l(\mathbf{r}_s)|^2 dS\right) \tag{5}$$

Eq.(5) is the popular relation for the prediction of perturbations by non-zero admittances. For a cylindrical cavity of the length $L$ and the inner radius $r$, Eq. (6) is reduced into

$$\frac{\Delta f_l}{f_l} \cong -\frac{1}{2}\text{Re}\left(\left(\frac{c \cdot j}{2\pi f_l \pi r^2 L \Lambda_l}\right)\iint_S y(\mathbf{r}_s, f_l)|\phi_l(\mathbf{r}_s)|^2 dS\right) \tag{6}$$

Eq. (6) shows that the perturbatroins are linearly dependent on cylinder length and integral of non-zero admittances. This linearity shapes the deletion mechanism for some types of perturbations by differential procedure. Suppose Cavity 1 and 2 of length $L_1$ and $L_2$ ($L_1 > L_2$), the perturbed resonant frequency $f_{l1}$ and $f_{l2}$ and the unperturbed resonant frequency $f_{l1}^0$ and $f_{l2}^0$. Imperfections are generally non-interactive so that the yielding perturbations on resonances are linearly additive. According to Eq. (6), unperturbed resonant frequencies can be extracted from perturbed resonant frequencies by:

$$f_{l1} \cong f_{l1}^0\left(1+\frac{\Delta f_1}{f_{l1}^0}\right) = \frac{l_1 c}{2L_1}\left(1+\frac{\Delta f_1}{f_{l1}^0}\right) \tag{7}$$

and

$$f_{l2} \cong f_{l2}^0\left(1+\frac{\Delta f_2}{f_{l2}^0}\right) = \frac{l_2 c}{2L_2}\left(1+\frac{\Delta f_2}{f_{l2}^0}\right) \tag{8}$$

in which, $\Delta f_1 = f_{l1} - f_{l1}^0$ and $\Delta f_2 = f_{l2} - f_{l2}^0$ denote the lumped sums of perturbations for Cavity 1 and 2, respectively; $l_1$ and $l_2$ stand for the eigenvalues of resonant modes for Cavity 1 and 2.

Having the identical inner radius $r$ and the length ratio of $L_1:L_2=2:1$ for the pair cavities, the resonant standing waves in the cavities will have equal frequencies $f_{l1}^0 = f_{l2}^0 = f^0$ on the condition of $l_1=2l_2$. Substituting Eq. (7) into Eq. (8) yields the speed of sound $c$:

$$c = \frac{2 f_{l1} f_{l2} \Delta L_{12}}{l_1 f_{l2} - l_2 f_{l1}} \left(1 - \frac{\delta L_{12}}{\Delta L_{12}}\right) \tag{9}$$

in which,

$$\frac{\delta L_{12}}{\Delta L_{12}} \cong \frac{L_1}{f^0 \Delta L_{12}} \left(\Delta f_1(f_{l1}, L_1) - \frac{1}{2} \Delta f_2(f_{l2}, L_2)\right) + O\left(\left(\frac{\Delta f_{1,2}}{f^0}\right)^2\right) \tag{10}$$

and $\Delta L_{12} = L_1 - L_2$.

$\delta L_{12}/\Delta L_{12}$ denotes the error for the calculation of $c$. Neglect the 2nd order term in Eq. (10). Substituting Eq. (6) into Eq. (10) yields

$$\frac{\delta L_{12}}{\Delta L_{12}} \cong \frac{c}{4 f^0 \Delta L_{12} (\pi r)^2} \mathrm{Re}\left(\frac{j}{\Lambda_{l2}} \iint_S y(\mathbf{r}_s, f_{l2}) |\phi_{l2}(\mathbf{r}_s)|^2 dS - \frac{j}{\Lambda_{l1}} \iint_S y(\mathbf{r}_s, f_{l1}) |\phi_{l1}(\mathbf{r}_s)|^2 dS\right) \tag{11}$$

As stated above, velocity potentials $\phi_{l1}$ and $\phi_{l2}$ depend on the ratio of the axial coordinate of $\mathbf{r}_s$ to the cavity length. $\phi_{l1}$ and $\phi_{l2}$ as well as $\Lambda_{l1}$ and $\Lambda_{l2}$ will be equal if the ratios are equal for the pair cavities. On condition of $L_1:L_2=2:1$ and $l_1=2l_2$, equal resonant frequency is yield in both resonators. The value of $\delta L_{12}/\Delta L_{12}$ approximates zero if the above conditions are satisfied and the area $S$ on which each admittance acts is equal. This comes out the deletion mechanism that some perturbations will be deleted in principle without depending on any correction calculation if the conditions for the differential procedure are satisfied, for instance, the perturbations on endplates and those that their acting area $S$ is independent from the cavity length. This distinguishing character implies that even if we do not recognize some imperfections, we still know their perturbations will be deleted under the differential procedure. We will show the practical effect of the theoretical conclusion in the experimental section.

Except for the real part given in Eq. (6), the complex admittance results in the companion imaginary part of wave number $K_l$. The imaginary part is characterized by the half-widths of resonances that can be reduced from Eq. (4). Then, the excess half-widhts of resonances the difference between the measurement and the prediction of half-widths is given by:

$$\frac{g_l - g_{\text{Theory}}}{f_l} \cong \frac{g_l}{f_l} - \frac{1}{2} \text{Im} \left( \left( \frac{c \cdot j}{2\pi f_l V \Lambda_l} \right) \iint_S y(\mathbf{r}_S, f_l) |\phi_l(\mathbf{r}_S)|^2 \, dS \right) \quad (12)$$

In Eq. (12), $g_l$ denotes the half-width of the measured resonant profile; $g_{\text{Theory}}$ stands for the theoretical prediction of that half-width. The theoretical $g_{\text{Theory}}$ would match in principle the experimental $g_l$ so that the excess half-width ($g_l - g_{\text{Theory}}$) would be zero for ideal situation. Thus, the non-zero excess half-width actually measures the extent in which the theoretical prediction differs from the measurement. The imaginary and the real part are twin companion quantities in a complex admittance $y(\mathbf{r}_S, f_l)$. Accordingly, the examination of excess half-width is a crucial assessment of the validity of theoretical admittance model [3,4,5,6,7,8,9].

## 4. Experimental setup

The differential procedure was conducted with a pair of resonators that were made to satisfy the conditions requested by the differential procedure. The pair cylindrical resonators composed of the short cavity of 80 mm in length and the long cavity of 160 mm in length. The cavity of 80 mm in length had been used for the latest determination of $k_B$ [10]. Both cavities were made from the same block of bearing steel. They have equal mechanic properties, were made of the identical nominal inner and outer diameters. The disagreement between the lengths of the pair cavities is bound in 0.005 mm. The parallelism of the ends of each cylindrical cavity is in (0.5 ~ 0.7) μm. The roundness of the cylindrical cavities is bound in 0.01 mm. The endplates made of equal dimension and mechanical properties from the same block of optical quartz glass. The PZT transducers as well as the gas fill ducts were of identical dimensions for both cavities. They were detailed in the previous publication [10]. The endplates were clamped to each cylindrical shell by equal torque. Each cavity sat on an individual circular plate. Another circular plate was placed upon the upper endplate of the cavity. Both circular plates were tightly clamped to the cavity through four thin rods that were connected through springs to the upper flange endplate of the pressure vessel. Thus, each cavity was separately and freely hung inside the same pressure vessel. The used gas manifold was indifferently applied. A T – branch union was applied upstream to the gas fill ducts of both cavities. As a result, the gas pressures inside the pair cavities were in equilibrium. The gas fill duct of each cavity has equal inner diameter and length.

Like what we described in Ref. [10], each cavity accommodates a pairs of capsule platinum resistance thermometers, Fluke Hart 5686-001-B 25 Ω. Each thermometer was enclosed with a copper sleeve, which has one sealed end and the length of 100 mm, the outer diameter of 10 mm, the inner diameter of 5.6 mm, and the sensing head of 40 mm long. Each thermometer was covered with thermally conducting grease and inserted through the open end of a sleeve. Then, a cap was screwed onto the sleeve to hold the thermometer against the inner wall and blind end of the sleeve. The four leads from each thermometer passing through a thin stainless-steel tube of the inner diameter of 2.13 mm and the length of 500 mm were welded to a sealed feed-through. The sleeve assembly was evacuated, purged with pure argon gas, and finally filled with pure argon gas to 110 kPa and sealed with a valve. The sleeve assembling was detailed in Ref. [10]. The thermometers were assembled into the sleeves and sealed before the calibrations in the triple point of water (TPW). The calibrations were conducted before and after the experiment. The assemblies were maintained entire through the whole acoustic measurements and the calibrations. Therefore, the sleeves allowed the thermometers maintaining a constant thermal resistance inside the TPW cell and the wells inside the cavity shells. The sleeves also protected the thermometers from the hydrostatic pressure during the acoustic measurements. Both pairs of thermometers demonstrated the identical performances reported in Ref. [10]. The differences between the averages of the thermometers of each pair were within ± 0.1 mK. The differences account for the temperature inhomogeneity inside each cavity. We averaged the reads of each pair thermometers through an entire measurement for a mode at an isobar. We observed the two averages differing within ± 0.3 mK. The differences indicate the temperature inhomogeneity inside the pressure vessel. The differences between the averages of the initial and final calibrations of each thermometer were bound in ± 0.02 mK, which stands for the long term stability of thermometers. The above specifications are equivalent to that reported in Ref. [10]. A number of bilateral comparisons between the reference cells of NIM[1] and those of INRIM[2] and PTB[3] have shown good agreement that the differences are bound in ± 0.03 mK. The effect of H and O isotopic compositions has been corrected for all those TPW cells. On considering the above messages, we estimate the uncertainty of 0.05 mK for the lumped mean value of the NIM's TPW reference cells.

The lengths of the cavities were measured by the two-color laser interferometry. The interferometer setup and the measurement procedure were indifferent from those reported in the latest experiment [10]. The laser beam of each color coming out from one laser source was divided into two branch beams of equal intensity. Each beam was led to the respective cavity through a separate and symmetric path. Each cavity had its own probing system for interference patterns. Therefore, the length of each cavity was measured independently by the identical procedure detailed in Ref. [10]. The length measurement uncertainty was in $0.28 \times 10^{-6}$ for the length of 160 mm and in $0.81 \times 10^{-6}$ for the length of 80 mm.

---

[1] National Institute of Metrology, China

[2] Istituto Nazionale di Ricerca Metrologica, Italy

[3] Physikalisch – Technische Bundesanstalt, Germany

The BIP Plus argon from the original container reported in Ref. [10] was applied for the experiment. As stated above, the BIP Plus argon sample has been analyzed in KLPRR CAS for the isotopic abundance ratios. We finished the comparison of its molar mass with the BIP argon sample that has been analyzed with the primary standard in KRISS. The comparison agrees well with the analytical result. Thus, the molar mass reported in Ref. [10] will be adopted for the new experiment with the differential procedure.

During measurements, each cavity is tested in different resonant mode in order to avoid possible coupling between resonances. Two parallel and independent acoustic signaling and detection systems were applied for the pair cavities. The electronic devices adopt the same models described in Ref. [10] for their specifications.

## 5. Experiment and data
### 5.1 Modes for the differential procedure

The natural vibrations of each cavity were inspected respectively in vacuum. The frequency spectral showed significantly increasing amplitudes at frequencies above 7.2 kHz for both cylindrical cavities. Figure 3 pictures the frequency spectral of each cavity. The spectrum near the natural vibrations of either cavity shows sharp peaks indicating intense vibration energies. If an acoustic resonance is formed near the natural vibration of cavity, they will have strong interaction. Thus, accurate resonant measurements have to be kept distant away from the natural vibrations of cavities. According to the inspections, the differential procedure has to be operated at modes below the criteria frequency 7.2 kHz. Besides, the free recoils of resonator ensemble are longitudinally symmetric for even modes so to cause no disturbance on resonances. In contrast, they are longitudinally asymmetric for odd modes so to yield disturbances. Free recoil is extremely hard to be quantitatively characterized because of poor knowledge of the mechanical forces connecting a cavity and its supporting components. If the even and the odd mode appear together in Eq. (11), an error cannot be deleted by the differential deletion mechanism. Accordingly, maintaining $l_1$ and $l_2$ both even integrals yields no free recoil disturbance with the differential procedure. By combining the considerations on the free recoil and the criterion of natural vibration of cavities, the compatible modes for the differential procedure is (200) for the short cylindrical cavity and (400) for the long cylindrical cavity.

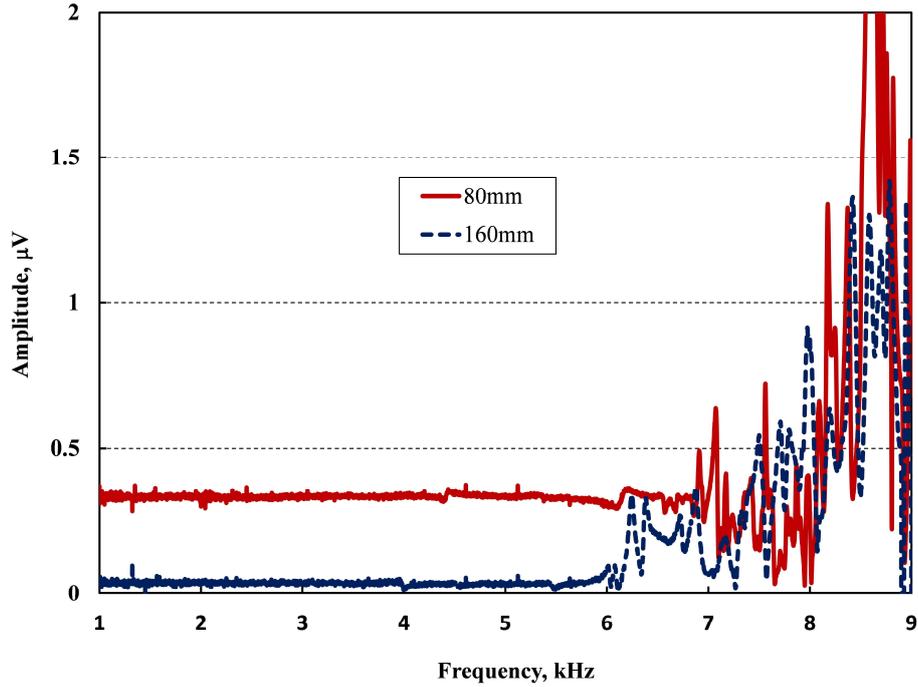

Figure 3 Spectral of the natural vibration of the pair cavities

Eq. (11) demonstrates that the differential procedure relies on the data of each single cavity. Thus, the differential procedure results relate to the reliable measurements of each single cavity. Accordingly, we examined the measurements for (200) of the short cavity and (400) of the long cavity to attest their measurements. The examination addresses the excess half-widths of resonant profiles and the resultant parameters of surface fitting such as the first and the second acoustic virial coefficients of each single cavity. At first, the excess half-widths for mode (200) of the short cavity and mode (400) of the long cavity is plotted in Figure 4. The excess half-widths of each cavity show the normal variation patterns that are similar to that of the latest experiment [10]. The upwarping at low pressures is considered due to low signal – to – noise of measurements in the pressure range.

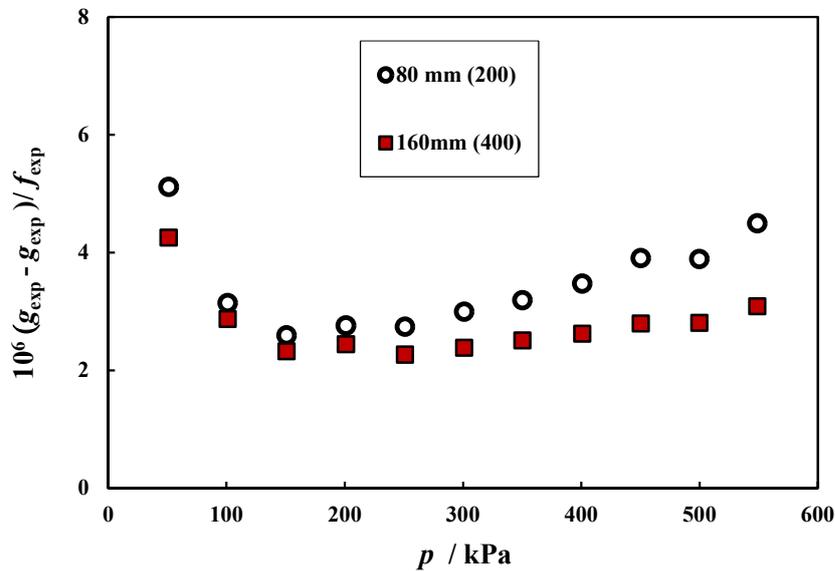

Figure 4 Comparison of the excess half-widths of mode (200) and (400)

Because the half-widths are due mostly to the thermal and viscous boundary layers, the leading pressure dependence of $(p/\text{kPa})^{1/2}$ and $(p/\text{kPa})^{-1/2}$ is expected. The data of each cavity consist of 282 measurements. Thus, the half-widths of each mode were fit separately with a function of the form:

$$g_{\text{fit}} = \frac{b_0}{(p/\text{kPa})^{1/2}} + b_1 + b_2 (p/\text{kPa})^{1/2} \qquad (13)$$

where $b_0 = 40.40763$, $b_1 = 2.9505\times10^{-2}$, and $b_2 = 2.450\times10^{-3}$ for the (200) mode and $b_0 = 34.14592$, $b_1 = 2.5065\times10^{-3}$, and $b_2 = 1.655\times10^{-3}$ for the (400) mode. We pictures the scattering of the measured half-widths with respect to the fitting values in Figure 5. We observe that the data are evenly scattered about the baseline with little or no systematic deviations present. The picture shows the increasing random errors with decreasing gas pressures.

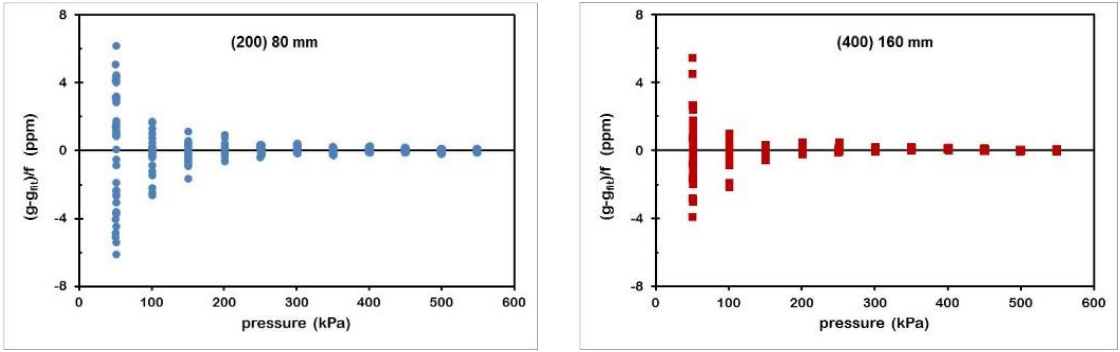

Figure 5 Scattering of half-widths of mode (200) and (400)

The resonant frequencies were fitted with a function of the form:

$$f_{\text{fit}}(T_w, p) = \frac{a_0}{(p/\text{kPa})^{1/2}} + a_1 + a_2 (p/\text{kPa}) + a_3 (p/\text{kPa})^2 \qquad (14)$$

where $a_0 = -39.18030$, $a_1 = 3849.6385$, $a_2 = 4.1643\times10^{-3}$, and $a_3 = 1.02491\times10^{-6}$ for the (200) mode and $a_0 = -33.14649$, $a_1 = 3849.7788$, $a_2 = 4.3712\times10^{-3}$, and $a_3 = 1.04840\times10^{-6}$ for the (400). The scattering of the data from the fitting values were plotted in Figure 6. Like the excess half-width data, the frequency deviations are evenly scattered about the baseline with little or no systematic deviations present.

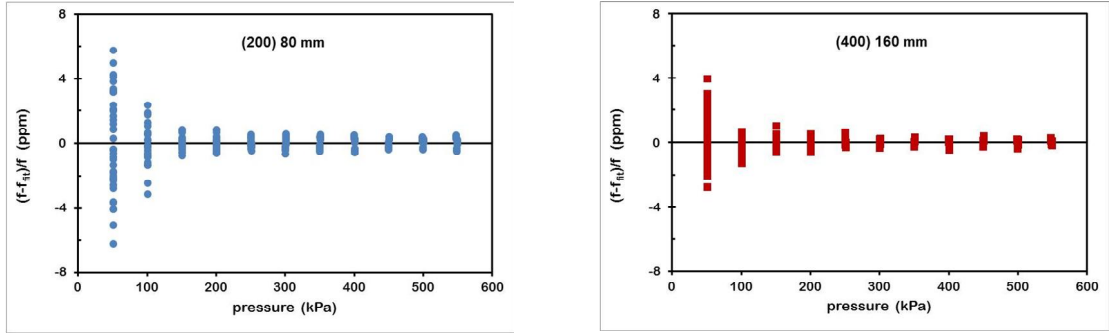

Figure 6 Scattering of resonant frequencies of mode (200) and (400)

We tabulated in Table 3 the parameters of the surface fittings due to Eq. (2) for the data of mode (200) and mode (400), and the references given by the AGT of a spherical cavity [3]. The parameter $A_0$ denotes the square speeds of sound of measurements at zero pressure. We observe $A_0$ at mode (200) of the 80 mm cavity is smaller than the reference in the fractional differences of $5.18 \times 10^{-6}$, and $A_0$ at (400) of the 160 mm cavity smaller than the reference in the fractional difference of $2.43 \times 10^{-6}$. The differences are compatible with the measurement uncertainty budgets of the previous measurements by the AGTs of single cavities [9,10]. The parameters $A_1$ and $A_2$ of each single cavity agree well the references.

Table 3 Comparison of the parameters of surface fitting

| parameter | unit | Mode (200) of 80 mm cavity | Mode (400) of 160 mm cavity | Reference [xx] |
|---|---|---|---|---|
| $A_0$ | m² s⁻² | 94755.69 | 94755.95 | 94756.18 |
| $10^4 A_1$ | m² s⁻² Pa⁻¹ | 2.2519 | 2.2452 | 2.2502 |
| $10^{11} A_2$ | m² s⁻² Pa⁻² | 5.287 | 5.277 | 5.321 |

In summary, the above examinations and comparisons attest the reliability of the parameters with the resonances of mode (200) of the 80 mm cavity and mode (400) of the 160 mm cavity.

**5.2 Application of the differential procedure**

In the above section, we have attributed the perturbations with AGTs of single fixed-length cavities to four types of imperfections. The perturbations are summed no interactions. Therefore, they are linearly additive as:

$$-\Delta f = f_l^0 - f_l = -\Delta f_b - \Delta f_d - \Delta f_{sh} - \Delta f_{tr} \tag{15}$$

where, the subscript "b" stands for boundary layers; "d" stands for gas fill duct; "sh" stands for shell motions (including endplate bending and cavity assembly free recoil); "tr" stands for transducers. The viscosity of gas causes the formation of viscous and thermal boundary layers on cavity shell, and the thermal boundary layer on each endplate. Thus, the perturbations yielded by boundary layers are summed as:

$$\Delta f_b = \Delta f_{th,\,endplate} + \Delta f_{th,shell} + \Delta f_{v,\,shell} \tag{16}$$

in which, the subscript "b" stands for boundary layers, "th" stands for thermal boundary layer and "v" accounts for viscous boundary layer, "endplate" and "shell" denote at endplate and shell of cylindrical cavity.

Likewise, the perturbations yielded by shell motions are summed by $\Delta f_{sh}$:

$$\Delta f_{sh} = \Delta f_{sh,\,radial} + \Delta f_{sh,\,axial} + \Delta f_{sh,\,end} + \Delta f_{sh,\,recoil} \tag{17}$$

in which, the subscript "radial" and "axial" stands for the shell motions in the axial and radial directions, respectively; "end" stands for the bending at the endplates; "recoil" stands for the free recoil of cavity assembly.

According to the dependency of cavity length, the perturbations are regrouped into two sets. Eq. (6) shows that the admittances acting on area $S$ that is independent of the shell area of cylindrical cavity will yield perturbations inversely proportional to cavity length. Such perturbations are catalogued into Group I, they are:

$$\left(\Delta f_{\text{th, endplate}} + \Delta f_{\text{sh, endplate}} + \Delta f_{\text{d}} + \Delta f_{\text{tr}}\right) \propto \frac{1}{L} \tag{18}$$

Group II is the set of perturbations independent of cavity length. Such perturbations act on area $S$ that is dependent on the shell area of cylindrical cavity. Such perturbations caused by the viscous and thermal boundary layer on the cylindrical shell are the typical elements of Group II. Likewise, the perturbations caused by the radial and longitudinal motions of cylindrical shell belong to Group II.

On the condition of $L_1=2L_2$, the identical inner radius $r_1=r_2$, and $l_1=2l_2$, Eq. (10) is reduced into

$$\delta L_{12} = \frac{l_1 c}{2 f^0} \left(\frac{\Delta f_1}{f_1} - \frac{1}{2}\frac{\Delta f_2}{f_2}\right) \cong \frac{l_1 c}{2 f^0 f_1}\left(\Delta f_1 - \frac{1}{2}\Delta f_2\right) \tag{19}$$

According to Eq. (6), the perturbations in Group I have the relation, $2\Delta f_1 \cong \Delta f_2$ in the ideal situation. Thus, $\delta L_{12}$ approaches zero. The deletion mechanism is effective for the elements in that group, but is ineffective for elements in Group II because there $\Delta f_1 \cong \Delta f_2$. In later situation, the operation of the differential procedure yields the remains of perturbations:

$$\Delta f_1 - \frac{1}{2}\Delta f_2 = \Delta f_{\text{th, shell, 1-2}} + \Delta f_{\text{v, shell, 1-2}} + \Delta f_{\text{radial, axial, 1-2}} \qquad (20)$$

in which, the subscript "1-2" stands for the operation of the differential procedure on the pair cavities; $\Delta f_{\text{th,shell,1-2}}$ is the remains of the perturbation of the thermal boundary layer on cylindrical shells; $\Delta f_{\text{v,shell,1-2}}$ is the remains of the perturbation of the viscous boundary layer on cylindrical shells; $\Delta f_{\text{radial,axial,1-2}}$ is the remains of the perturbations of the longitudinal and radial motions of cylindrical shells. The perturbations of boundary layers largely predominate over all other perturbations, nevertheless they are well characterized by the theory of the first order approximation and the further second order approximation [17,23,24]. The viscosity accuracy of sample gas dominates the prediction of boundary layer perturbations. For argon, the excellent agreement within ± 0.04 % at TPW has been achieved among the state-of-the-art results given by the *ab initio* calculations [25,26] and the experiments [27,28,29,30,31,32]. The achieved accuracy for the argon viscosity at TPW is expect to improve the correction of boundary layer perturbations to the level bound in $0.1 \times 10^{-6}$.

We compared the square speeds of sound obtained from the differential procedure and the reference data measured by the AGT of spherical resonator [3]. In order to examine the deletion effect of the differential procedure, we composed three situations denoted by Opt.1, Opt.2 and Opt.3, that are correspondent with three different degrees of perturbation corrections. Opt.1 denotes the results of the differential procedure that the data of each single cavity were corrected from all the perturbations in Group I and II; Opt2 denotes the results of the differential procedure that the data of each single cavity were corrected from all the perturbations of the boundary layers formed on the cylindrical shell and the pair endplates; Opt.3 denotes the results of the differential procedure that the data of each single cavity were only corrected from the perturbations of the boundary layers formed on the cylindrical shell. The results of three options were shown together in Figure 7 (a) in which blue diamond solids, the red rectangular solids and the green triangle solids stand for results of Opt.1, Opt.2 and Opt.3, respectively. Figure 7 (a) shows that the deletion mechanism is significant for all three situations. All the differences converge to the point at zero pressure with the fractional difference of $0.01 \times 10^{-6}$. Recalling that the three cases level the different degrees of perturbation corrections, the excellent agreement of the three cases at zero pressure attests the above analysis of the deletion mechanism of the differential procedure. The value of $A_0$ (see the following paragraph about the surface fitting), the square speeds of sound at zero pressures, is in the fractional difference of $0.18 \times 10^{-6}$ above the reference. We observed the slight fractional differences increasing with increasing gas pressures, and approximate linearly vanishing at zero pressure. The phenomenon implies the remains of the perturbations of shell motions, such as those in Eq. (20). The differences slightly vary from situation to situation. The measurements of Opt.1 are smaller than the references in - $(0.5 \sim 7.1) \times 10^{-6}$ in the entire pressure range. The differences are slightly larger than those of Opt.2 and 3. The largest difference of $-7.1 \times 10^{-6}$ appears at the maximum pressure. The slight differences among Opt.1, 2 and 3 imply that the corrections of the perturbations of shell motions may be slightly improperly done. The measurements of Opt.2 and Opt.3 match closely. Their fractional differences bound in $0.1 \times 10^{-6}$ in the entire pressure range. The comparison attests the effectiveness of the deletion mechanism for the perturbations of the viscous and thermal boundary layers on the endplates of the pair cavities. The results of Opt. 2 and 3 get closer to the references in the whole pressure range. The measurements of Opt.2 and 3 differs from the references in - $(0.7\sim4.2)\times10^{-6}$ in the entire pressure range. The largest difference appears at the maximum pressure. Such differences imply that the differential procedure is without turning out significant remains of the perturbations caused by the radial and longitudinal motions of the cylindrical shells. Those remains consistently approximate vanishing at zero pressure.

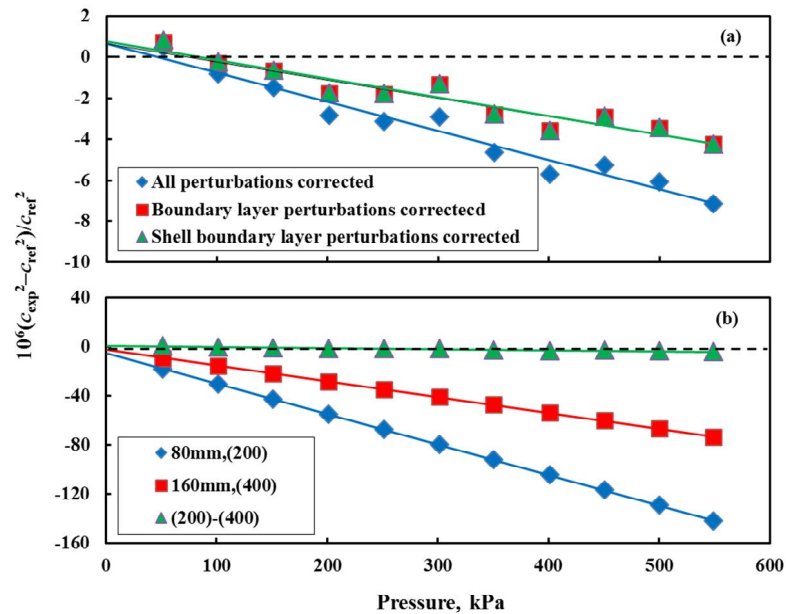

Figure 7 Comparison of square speeds of sound

Figure 7 (b) plots the comparison between the data of Opt.2 and the data of the single 80mm and 160 mm cavities of the pair (as stated above, all perturbations caused by the shell motions were not corrected from the data of each single cavity for Opt.2). The comparison was conducted over the entire pressure range of the experiment. Because of no correction of the perturbations of the shell motions, the data of each single cavity differ significantly from the references at high pressures. The differences appear in the typical character of perturbations of shell motions. They approximate vanishing at zero pressure. We observed that the differences with the 80 mm cavity are constantly in nearly two folds larger than those with the 160 mm cavity. This difference ratio actually projects the relation, $2\Delta f_1 \cong \Delta f_2$, for the perturbations in Group I. The data yielded from the differential procedure form a prominent contrast to the data of both single cavities. The contrast signifies the effectiveness of the deletion mechanism. The differences are of one order smaller than those with each single cavity in the entire pressure range.

Besides the square speeds of sound, the half-width of resonance is an important parameter for attesting the performance of acoustic procedures. The kinetic energy dissipation property of boundary layers, the viscous effect of bulk gas in cavity and ducts is the cause for the formation of half-widths of acoustic resonant spectra. The total half-width is counted in:

$$g = g_b + g_{bulk} + g_d \tag{21}$$

where the subscript "bulk", "b" and "d" stands for bulk gas, boundary layer and duct, respectively. The excess half-widths of the differential procedure are given by:

$$\left(\frac{g_l - g_{Theory}}{f_l}\right)_1 - \left(\frac{g_l - g_{Theory}}{f_l}\right)_2$$

The theoretical predictions of the half-widths of each single cavity were calculated according to Eq. (12). They were plotted in Figure 8. The maximum excess half-width of $2.8 \times 10^{-6}$ appears at the maximum pressure. The excess half-widths tend obviously to zero with decreasing gas pressures. The plot shows a significant improvement comparing with the diagrams of excess half-widths of AGTs of single cavity [9,10] and those presented in Figure 4. The examination proves on the other hand of the effectiveness of the differential procedure.

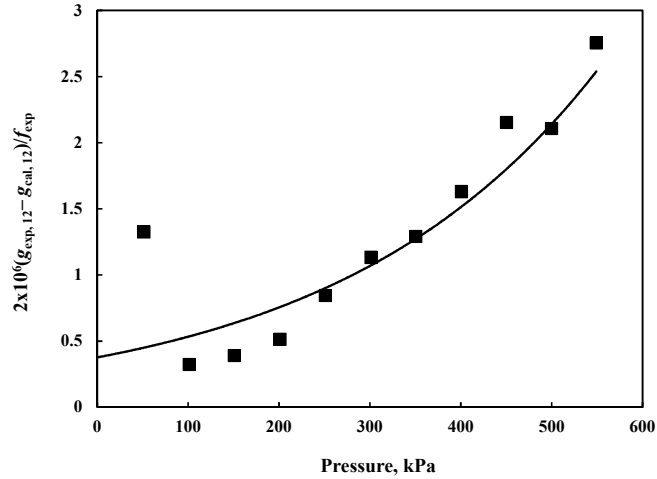

Figure 8 Excess half-widths of the resonances with the differential procedure

According to Eq. (9) and Eq. (19), the speeds of sound $c_{exp}$ ($T_{TPW}$, $p$) are calculated with the differential procedure based on the measurements of each single cavity of the pair. By knowing $c_{exp}$ ($T_{TPW}$, $p$), the acoustic virial relation, Eq. (2), is formed by the surface fitting of the 4-parameter function of pressure $p$. Eq. (2) is rewritten as:

$$c_{exp}^2 - A_3 p^3 = A_0 + A_1 p + A_2 p^2 + A_{-1} p^{-1} \qquad (22)$$

In Eq. (22), $A_0 \equiv c_0^2 = \gamma_0 k_B T_{TPW}/M_{Ar}$ where $M_{Ar}$ stands for the molar mass of the BIP Plus argon sample used. When fitting Eq. (22) with $c_{exp}$ ($T_{TPW}$, $P$) given by the differential procedure, we equally weighted the values of $(c_{exp})^2$ at each pressure from 50 kPa to 550 kPa. In Figure 7 (a), we have observed excellent agreement of the fractional difference of $0.01 \times 10^{-6}$ among the values of $A_0$ of three options. Besides, there are excellent agreements among their second and third acoustic virial coefficients for the three situations that the fractional differences bound in ± 0.1 % and ± 0.01 %, respectively. Thus, we selected the second situation, Opt.2, for yielding the parameters of $A_0$, $A_1$ and $A_2$ according to Eq. (22). Those yielded parameters are tabulated in Table 4. We illustrate in Table 4 the references of those parameters [3]. Except for the good agreement of $A_0$ stated above, the values of $A_1$ and $A_2$ are in good agreement, respectively.

Table 4 Parameters of surface fitting the data of the differential procedure

| parameter | unit | Results of differential procedure | Measurements of spherical cavity [3] |
|---|---|---|---|
| $A_0$ | m² s⁻² | 94756.195(178) | 94756.178(65) |
| $10^4 A_1$ | m² s⁻² Pa⁻¹ | 2.2418(87) | 2.2502(35) |
| $10^{11} A_2$ | m² s⁻² Pa⁻² | 5.3435(1169) | 5.3210(62) |

The second acoustic virial coefficient $\beta_\alpha$ is reduced from the parameter, $A_1$, due to the relation $\beta_\alpha = (M_{Ar}/\gamma_0)A_1$ where $\gamma_0=5/3$ for monoatomic gases; the third acoustic virial coefficient $\gamma_\alpha$ is reduced from the parameter, $A_2$, due to $\gamma_\alpha = (M_{Ar}/\gamma_0)A_2$. We tabulate the state-of-the-art experimental and theoretical results of the second and third acoustic virial coefficients [2,3,8,23,24,25,33,34,35,36,37,38] in Table 5. Figure 9 (a) and (b) picture respectively the comparisons of those data. The plot in Figure 9 (a) shows the new measurement agree well with the state-of-the-art results for the second acoustic virial coefficients. Because those results come out from very independent sources, the good agreement attests the proper performance of the AGTs and the theoretical models. Figure 9 (b) shows that the third acoustic virial coefficients scatter larger than those of the second acoustic virial coefficients. Nevertheless, the value given by this work is closed to the one given by the AGT of spherical cavity [3]. In summary, the good agreement among the high order parameters adds the evidence on the performance of the differential procedure.

Table 5 Data of the second and third acoustic virial coefficients

| Authors | Year | $\beta_a$ cm$^3$·mol$^{-1}$ | $\gamma_a$ cm$^6$·mol$^{-2}$ |
|---|---|---|---|
| Colclough et al. [2] | 1979 | 4.842±0.072 | 1668±89 |
| Ewing et al. [34] | 1986 | 5.240±0.062 | 1683±189 |
| Moldover et al. [3] | 1988 | 5.401±0.010 | 1275±15 |
| Ewing et al. [35] | 1989 | 5.290±0.05 | 3000±200 |
| Ewing and Trusler [36] | 1992 | 5.575±0.08 | 2750±300 |
| Ewing and Goodwin [37] | 1992 | 5.452±0.009 | 1217±4 |
| Estrada-Alexanders and Trusler [38] | 1995 | 5.51±0.04 | 2658±80 |
| Moldover et al. [23] | 1999 | 5.401±0.020 | 1275±30 |
| Benedetto et al. [24] | 2004 | 5.344±0.035 | 1393±66 |
| Pitre et al. [33] | 2006 | 5.416 | - |
| Sutton et al. [8] | 2010 | 5.387 | - |
| Vogel et al. [25] | 2010 | 5.443 | - |
| Lin et al. [10] | 2013 | 5.465±0.022 | 1283±15 |
| This work | 2014 | 5.373±0.011 | 1300±15 |

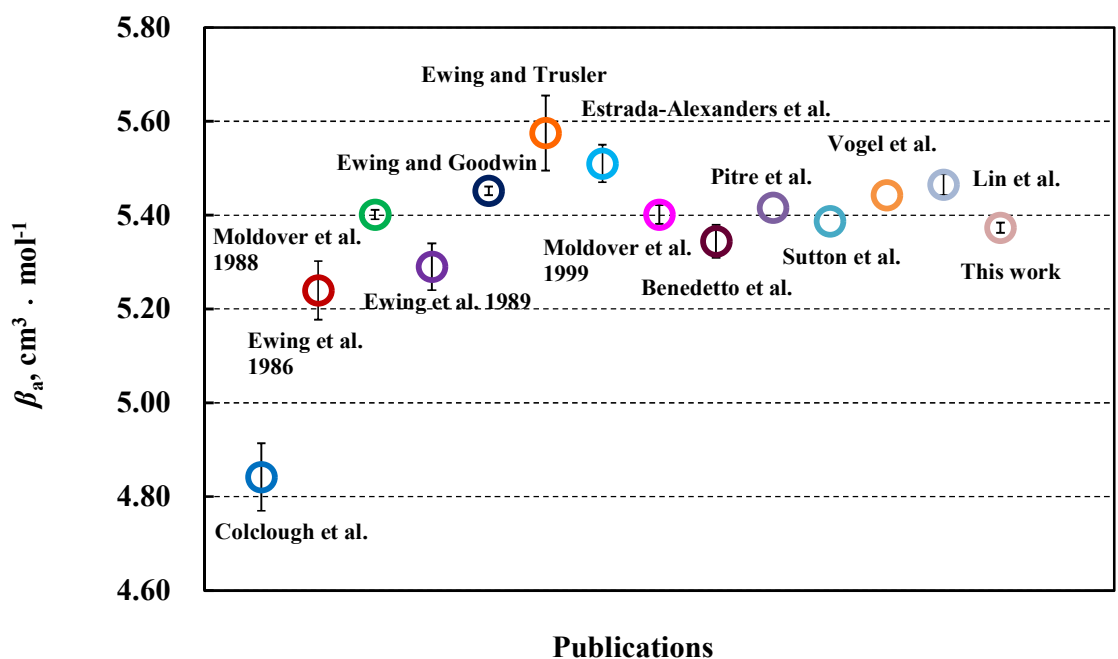

Figure 9 (a) Comparison of the second acoustic virial coefficients

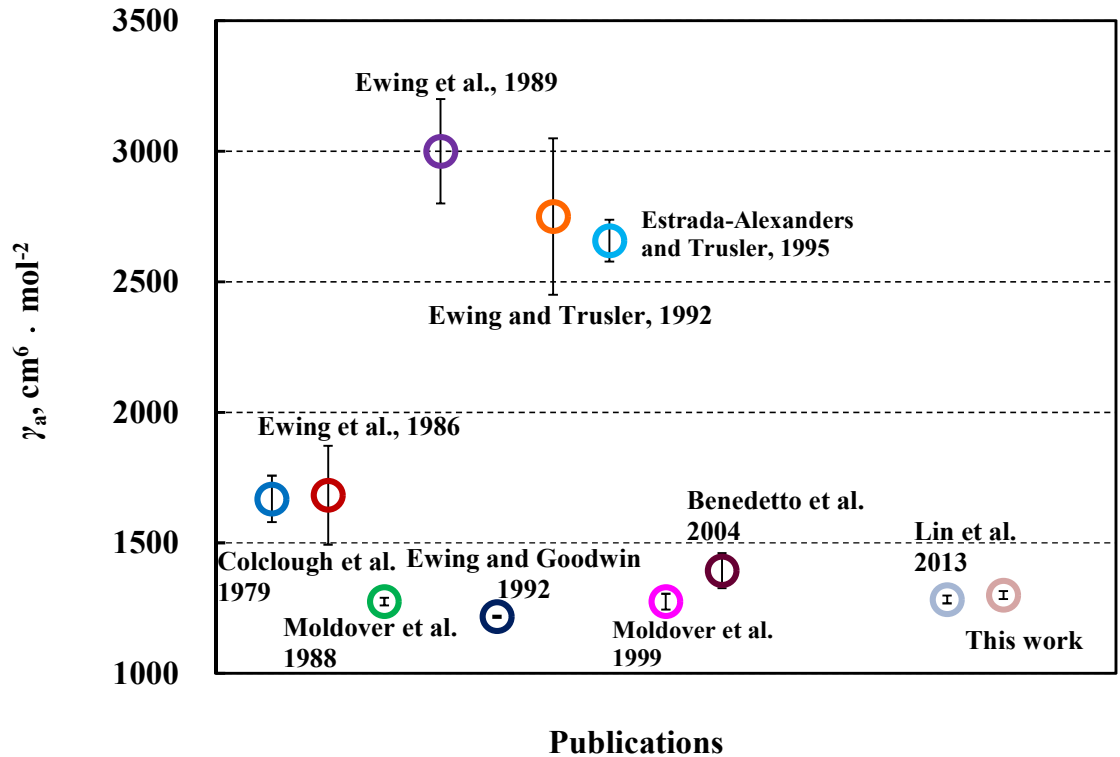

Figure 9 (b) Comparison of the third acoustic virial coefficients

According to Eq. (1), the Boltzmann constant $k_B$ is reduced from $c_0^2$ on condition of knowing the molar mass $M_{Ar}$ and the triple point of water $T_{TPW}$. The differential procedure yields the determination of $k_B$ equal to $1.380\,650\,6\times10^{-23}$ J K$^{-1}$ with the relative standard uncertainty of $2.3\times10^{-6}$. The new value of $k_B$ is above the adjusted value of CODATA 2010 [11] in the fractional difference of $1.3\times10^{-6}$. The resultant gas constant is $R = k_B N_A = 8.314\,473\,0$ J·mol$^{-1}$·K$^{-1}$ with the same the relative standard uncertainty. We plotted in Figure 10 the state-of-the-art determinations of $k_B$ [2,3,4,5,6,7,8,9,10,39,40] in the fractional differences from the adjusted value given by CODATA 2010.

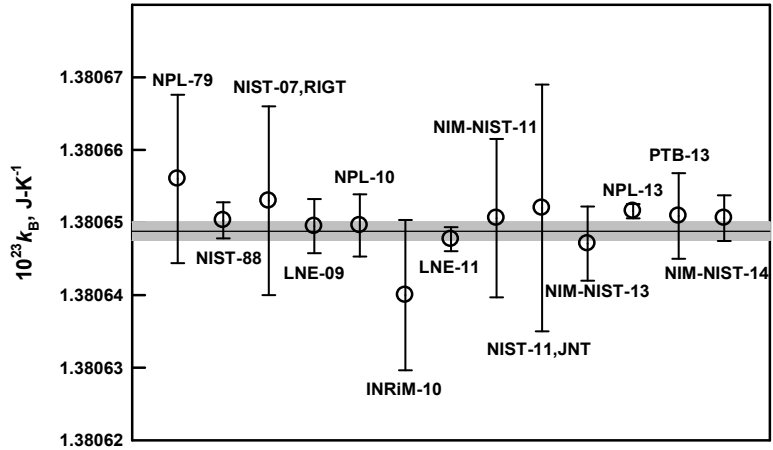

Figure 10 Comparison of the state-of-the-art determinations of $k_B$ relative to the recommended value given by CODATA 2010 [11]

### 5.3 Uncertainty budget

In summary, the operation of the differential procedure bases on the calculation of the measurements of the pair of single cylindrical cavities according to the differential principle stated by Eq. (9) and (10). The measurements are wholly done separately on each single cavity of the pair. Thus, the uncertainty for the determination of $k_B$ by the differential procedure has the contributions highly similar to those of AGT of single cavity. The uncertainty budget composes of the uncertainty contributions by: the measurement of $T_{TPW}$, the determination of the molar mass of Ar, the measurement of the lengths of both cavities, the curve fittings of both resonant profiles, the correction of the perturbation of the boundary layers formed on cylindrical shell, and the surface fitting. We illustrated those contributions in Table 6. The combined relative standard uncertainty, $2.3 \times 10^{-6}$ for the new determination of $k_B$, is calculated depending on those contributions. We observed the random error with the surface fitting for $A_0$ comes out the largest uncertainty contribution. We attribute that term to the low signal – to – noise acoustic response at low pressures.

Table 6 Uncertainty budget

| Uncertainty source | $10^6 \times u_r(k_B)$ |
|---|---|
| 1. Gas temperature measurement | |
|    Thermometer calibration | 0.36 |
|    Temperature gradient | 0.17 |
|    TPW realized by the reference cell | 0.18 |
| 2. Avogadro constant, | 0.05 |
| 3. Molar mass | |
|    Abundance of noble gas impurities | 0.03 |
|    Isotopic abundance ratios | 0.77 |
| 4. Length measurement | |
|    Length of the long cylinder | 0.28 |
|    Length of the short cylinder | 0.81 |
| 5. Zero-pressure limit of corrected frequencies | |
|    Boundary layer corrections | 0.40 |
|    Random error in $A_0$ | 1.87 |
|    dispersion | 0.07 |
|    Combined uncertainty | 2.3 |

## 6. Discussion and summary

We have reported the latest studies for the determination of $k_B$ by the AGT of fixed-length cylindrical cavity. First, we present the results of the study of acoustic comparison for molar masses of sample gases. The study covers the sample of the BIP argon (labeled as C) analyzed by the GC-MS in KRISS, the sample of the BIP Plus (labeled as AB) analyzed by the GC-MS in KLRPP CAS and a BIP Plus sample (labeled as D) from an arbitrary cylinder. The ratio of the molar masses of a pair samples is available by comparing their square speeds of sound on an isotherm and isobar. Thus, the study provides an independent inspection of the linkage of results between KRISS and KLRPP CAS. The ratio of the molar masses between Sample C and AB measured by the acoustic method agree well with that given by the GC-MS in KRISS and KLRPP CAS. The ratio between Sample C and D is given by the study. The authors have sent Sample D to KRISS for a new analysis by the GC-MS, and to Laboratoire Commun de Metrologie LNE-Cnam for an acoustic comparison by the AGT of quasi spherical cavity. Thus, the results of the current study provide the useful data for the comparison with the results from some independent studies.

Second, we present in detail the analysis and experiment for the AGT of fixed – length cylindrical cavity operating in the novel differential – cylindrical procedure. The analytical and experimental results consistently show the deletion mechanism for the perturbations of Group I. The new mechanism is highlighted by the fact that it is free from correction of Group I perturbations. This character distinguishes the new AGT from the classical ones of single fixed – length cylindrical cavities. The theoretical model of the new method hints that the perturbations of unknown causes can be deleted by the differential mechanism if the perturbations are sorted into Group I. The analysis shows that the new method is ineffective for the perturbations of Group II. The predominant elements of Group II are the perturbations caused by the thermal and viscous boundary layers formed on the cylindrical shells and endplates. The comprehensive experimental and theoretical studies have attested that such perturbations can be well corrected by the theoretical prediction due to the accurate knowledge of viscosities of sample gases. The comparison of the data given by Opt.2 and 3 have shown that the remains in Group II yield extremely minor effect on the resultant acoustic parameters of the surface fitting. The second and third acoustic virial coefficients yield by the new procedure agree quite well with the data of the state-of-the-art theoretical and experimental studies. The excess half-widths approximate vanishing at zero pressure, showing the significant improvement against the data of AGTs of single fixed-length cavities. In summary, all the acoustic parameters examined point out that the new procedure behaves properly.

The Boltzmann constant $k_B$ is redetermined to be 1.380 650 6×10$^{-23}$ J K$^{-1}$ by the new procedure that operates in the mechanism very different from those led to the CODATA 2010 value of $k_B$. Our result differs from the CODATA 2010 value of $k_B$ in the fractional difference of (1.3±2.3)×10$^{-6}$. The resultant gas constant $R$ is equal to 8.314 473 0 J·mol$^{-1}$·K$^{-1}$ with the same uncertainty for the redetermination of $k_B$.

**Acknowledgements**

We thank the colleagues, Dr. Laurent Pitre, Dr. Michael de Podesta, Dr. Roberto Gavioso and Mr. Robin Underwood for their helpful discussions and encouragement. This work was supported by the National Natural Science Foundation of China (Nos. 51476153, 51276175 and 51106143).